\documentclass[12pt]{article}
\usepackage{epsf, amssymb,amsmath,dsfont}
\usepackage{epsfig}
\usepackage{hyperref}

\makeatletter
\renewcommand\section{\@startsection {section}{1}{\z@}%
                                   {-3.5ex \@plus -1ex \@minus -.2ex}
                                   {2.3ex \@plus.2ex}%
                                   {\normalfont\large\bfseries}}

\renewcommand\subsection{\@startsection{subsection}{2}{\z@}%
                                     {-3.25ex\@plus -1ex \@minus -.2ex}%
                                     {1.5ex \@plus .2ex}%
                                     {\normalfont\bfseries}}

\makeatother

\newcommand{\bea}{\begin{eqnarray}}
\newcommand{\eea}{\end{eqnarray}}
\newcommand{\be}{\begin{equation}}
\newcommand{\ee}{\end{equation}}
\newcommand{\bem}{\begin{pmatrix}}
\newcommand{\eem}{\end{pmatrix}}
\newcommand{\bl}{\begin{align}}
\newcommand{\el}{\end{align}}

\begin{document}
\begin{center} 
${}$
\thispagestyle{empty}

\vskip 3cm {\LARGE {\bf On the Origin of Gravity \\[4mm] and the Laws of Newton} 
}
\vskip 1.25 cm  { Erik Verlinde\footnote{
e.p.verlinde@uva.nl} }\\
{\vskip 0.5cm  Institute for Theoretical Physics\\ University of Amsterdam\\
Valckenierstraat 65\\
1018 XE, Amsterdam\\
The Netherlands\\}

\vspace{2cm}

\begin{abstract}
\baselineskip=16pt
Starting from first principles and general assumptions Newton's law of gravitation is shown to arise naturally and unavoidably in a  theory in which space is emergent through a holographic scenario. Gravity is explained as an entropic force caused by changes in the information associated with the positions of material bodies.  A relativistic generalization of the presented arguments directly leads to the Einstein equations. When space is emergent even Newton's law of inertia needs to be explained. The equivalence principle leads us to conclude that it is actually this law of inertia whose origin is entropic.
\end{abstract}
\end{center}

\newpage

\section{Introduction}

Of all forces of Nature gravity is clearly the most universal. Gravity influences and is influenced by everything that carries an energy, and is intimately connected with the structure of space-time.
The universal nature of gravity is also demonstrated by the fact that its basic equations closely resemble the laws of thermodynamics and hydrodynamics\footnote{An incomplete list of references includes \cite{Bekenstein, Bardeen,  Hawking, Davies, Unruh, Damour, Jacobson}}.  So far, there has not been a clear explanation for this resemblance.

Gravity dominates at large distances, but is very weak at small scales. In fact, its basic laws have only been tested up to distances of the order of a millimeter.  Gravity is also considerably harder to combine with quantum mechanics than all the other forces. The quest for unification of gravity with these other forces of Nature, at a microscopic level, may therefore not be the right approach.  It is known to lead to many problems, paradoxes and puzzles. String theory has to a certain extent solved some of these, but not all. And we still have to figure out what the string theoretic solution teaches us. 

Many physicists believe that gravity, and space-time geometry are emergent.   Also string theory and its related developments have given several indications in this direction. 
Particularly important clues come from the AdS/CFT, or more generally, the open/closed string correspondence. This correspondence leads to a duality between theories that contain gravity and those that don't. It therfore provides evidence for the fact that gravity can emerge from a microscopic description that doesn't know about its existence. 

The universality of gravity suggests that its emergence should be understood from general principles that are independent of the specific details of the underlying microscopic theory.  
In this paper we will argue that the central notion needed to derive gravity is information.  More precisely, it is the amount of information associated with matter and its location, in whatever form the microscopic theory likes to have it,  measured in terms of entropy.  Changes in this entropy when matter is displaced leads to an entropic force, which as we will show takes the form of gravity.  Its origin therefore lies in the tendency of the microscopic theory to maximize its entropy. 

The most important assumption will be that the information associated with a part of space obeys the holographic principle \cite{thooft, susskind}. The strongest supporting evidence for the holographic principle comes from black hole physics \cite{Bekenstein, Hawking} and the AdS/CFT correspondence \cite{maldacena}. These theoretical developments indicate that at least part of the microscopic degrees of freedom can be represented holographically either on the boundary of space-time or on horizons. 

The concept of holography appears to be much more general, however.  For instance, in the AdS/CFT correspondence one can move the boundary inwards by exploiting a holographic version of the renormalization group. Similarly, in black hole physics there exist ideas that the information can be stored on stretched horizons. Furthermore, by thinking about accelerated observers, one can in principle locate holographic screens anywhere in space. In all these cases the emergence of the holographic direction is accompanied by redshifts, and related to some coarse graining procedure. If all these combined ideas are correct there should exist a general framework that describes how space emerges together with gravity.

Usually holography is studied in relativistic contexts. However, the gravitational force is also present in our daily non-relativistic world. The origin of gravity, whatever it is, should therefore also naturally explain why this force appears the way it does, and obeys Newton law of gravitation. In fact,  when space is emergent, also the other laws of Newton have to be re-derived, because standard  concepts like position, velocity, acceleration, mass and force are far from obvious. Hence, in such a setting the laws of mechanics have to appear alongside with space itself.  Even a basic concept like inertia
is not given, and needs to be explained again.

In this paper we present a holographic scenario for the emergence of space and address the origins of gravity and  inertia, which are connected by the equivalence principle. Starting from first principles, using only space independent concepts like energy, entropy and temperature,  it is shown that Newton's laws appear naturally and practically unavoidably.  Gravity is explained as an entropic force caused by a change in the amount of information associated with the positions of  bodies of matter.

 A crucial ingredient is that only a finite number of degrees of freedom are associated with a given spatial volume, as dictated by the holographic principle. The energy, that is equivalent to the matter, is distributed evenly over the degrees of freedom, and thus leads to a  temperature. The product of the temperature and the change in entropy due to the displacement of matter is shown to be equal to the work done by the gravitational force.  In this way Newton's law of gravity emerges in a surprisingly simple fashion. 
 
The holographic principle has not been easy to extract from the laws of Newton and Einstein, and is deeply hidden within them. Conversely, starting from holography, we find that these well known laws come out directly and unavoidably.   By reversing the logic that lead people from the laws of gravity to holography, we will obtain a much sharper and even simpler picture of what gravity is. For instance, it clarifies why gravity allows an action at a distance even when there in no mediating force field.  

The presented ideas are consistent with our knowledge of string theory, but if correct they should have important implications for this theory as well.  In particular, the description of gravity as being due to the exchange of closed strings can no longer be valid. In fact, it appears that strings have to be emergent too. 

 We start in section 2 with an exposition of the concept of entropic force.  Section 3 illustrates the main heuristic argument in a simple non relativistic setting. Its generalization to arbitrary matter distributions is explained in section 4. In section 5 we extend these results to the relativistic case, and derive the Einstein equations.  The conclusions are presented in section 7.

\begin{figure}[bbp]
\begin{center}
\vspace{-.5cm}
\includegraphics[scale=0.25]{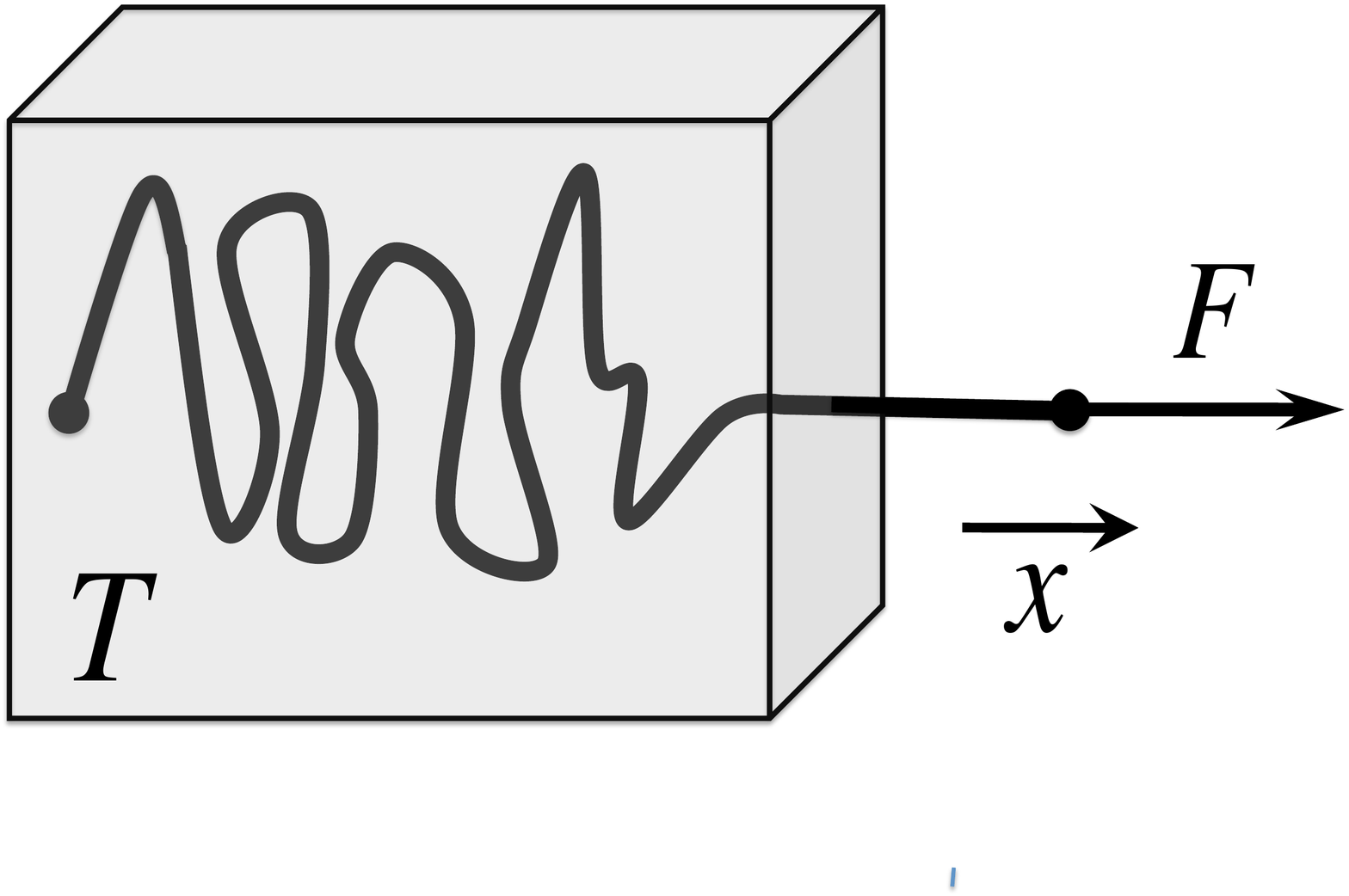}
\vspace{-1cm}
\caption{\small{A free jointed polymer is immersed in a heat bath with temperature $T$ and pulled out of its equilibrium state by an external force $F$. The entropic force points the other way.}}
\end{center}
\end{figure} 

\section{Entropic Force.}

An entropic force is an effective macroscopic force that originates in a system with many degrees of freedom by the statistical tendency to increase its entropy. The force equation is expressed in terms of entropy differences, and is independent of the details of the microscopic dynamics.  In particular, there is no fundamental field associated with an entropic force.  Entropic forces occur typically in macroscopic systems such as in colloid or bio-physics. Big colloid molecules suspended in an thermal environment of smaller particles, for instance, experience entropic forces due to excluded volume effects. Osmosis is another phenomenon driven by an entropic force.

Perhaps the best known example is the elasticity of a polymer molecule. A single polymer molecule can be modeled by joining together many monomers of fixed length, where each monomer can freely rotate around the points of attachment and direct itself in any spatial direction. Each of these configurations has the same energy. When the polymer molecule is immersed into a heat bath, it likes to put itself into a randomly coiled configuration since these are entropically favored. There are many more such configurations when the molecule is short compared to when it is stretched into an extended configuration.
The statistical tendency to return to a maximal entropy state translates into a macroscopic force, in this case the elastic force. 

By using tweezers one can pull the endpoints of the polymer apart, and bring it out of its equilibrium configuration by an external force $F$, as shown in figure 1.  For definiteness, we keep one end fixed, say at the origin, and move the other endpoint along the $x$-axis. The entropy equals
\be
S(E,x)=k_B\log \Omega(E,x)
\ee
where $k_B$ is Boltzman's constant and  $\Omega(E,x)$ denotes the volume of the configuration space for the entire system as a function of the total energy $E$  of the heat bath and  the position $x$ of the second endpoint. The $x$ dependence is entirely a configurational effect:  there is no microscopic contribution to the energy $E$  that depends on $x$.

In the canonical ensemble the force $F$ is introduced in the partition function\footnote{We like to thank B. Nienhuis and M. Shigemori for enlightning discussions on the following part.}
\be
Z(T,F)=\!\int\!\! dE dx\,\Omega(E,x)\, e^{-(E+Fx)/k_BT}.
\ee
as an external variable dual to the length $x$ of the polymer. The force $F$ required to keep the polymer at a fixed length $x$  for a given temperature $E$  can be deduced from the saddle point equations
\be
\label{twoequations}
{1\over T}= {\partial S\over \partial E},\qquad\quad
{F\over T}= {\partial S\over \partial x}.
\ee
By the balance of forces, the external force $F$ should be equal to the entropic force, that tries to restore the polymer to its equilibrium position. An entropic force is recognized by the facts that it points in the direction of increasing entropy, and, secondly, that it is proportional to the temperature. 
For the polymer the force can be shown to obey Hooke's law
$$
{F}_{polymer} \sim - {\rm const}\cdot\, k_BT\, \langle x\rangle.
$$
This example makes clear that  at a macroscopic level an entropic force can be conservative, at least when the temperature is kept constant. The corresponding potential has no microscopic meaning, however, and is emergent. 

It is interesting to study the energy and entropy balance when one gradually lets the polymer return to its equilibrium position, while allowing the force to perform work on an external system. By energy conservation, this work must be equal to the energy that has been extracted from the heat bath. The entropy of the heat bath will hence be reduced. For an infinite heat bath this would be by the same amount as the increase in entropy of the polymer. Hence, in this situation the total entropy will remain constant.    

This can be studied in more detail in the micro-canonical ensemble, because it takes the total energy into account  including that of the heat bath. To determine the entropic force, one again introduces an external force $F$ and examines the balance of forces. Specifically, one considers the micro canonical ensemble given by $\Omega(E+Fx,x)$, and imposes that the entropy is extremal. This gives
\be
\label{extremal1}
{d\over dx} S(E\!+\!Fx,\,x)=0
\ee 
One easily verifies that this leads to the same equations (\ref{twoequations}). However, it illustrates that microcanonically the temperature is in general position dependent, and the force also energy dependent.  The term $Fx$ can be viewed as the energy that was put in to the system by pulling the polymer out of its equilibrium position. This equation tells us therefore that the total energy gets reduced when the polymer is slowly allowed to return to its equilibrium position, but that the entropy in first approximation stays the same. Our aim in the following sections is to argue that gravity is also an entropic force, and that the same kind of reasonings apply to it with only slight modifications.

 \section{Emergence of the Laws of Newton}
 Space is in the first place a device introduced to describe the positions and movements of particles. Space is therefore literally just a storage space for information. This information is naturally associated with matter.  
Given that the maximal allowed information is finite for each part of space, it is impossible to localize a particle with infinite precision at a point of a continuum space. In fact, points and coordinates arise as derived concepts.  One could assume that information is stored in points of a discretized space (like in a lattice model). But if all the associated information would be without duplication, one would not obtain a holographic description. In fact, one would not recover gravity. 

Thus we are going to assume that information is stored on surfaces, or screens. Screens separate points, and in this way are the natural place to store information about particles that move from one side to the other. Thus we imagine that this information about the location particles is stored in discrete bits on the screens. The dynamics on each screen is given by some unknown rules, which can be thought of as a way of processing the information that is stored on it.  Hence, it does not have to be given by a local field theory, or anything familiar. The microscopic details are irrelevant for us. 

Let us also assume that like in AdS/CFT,  there is one special direction corresponding to scale or a coarse graining variable of the microscopic theory.  This is the direction in which space is emergent. So the screens that store the information are like stretched horizons. On one side there is space, on the other side nothing yet. We will assume that the microscopic theory has a well defined notion of time, and its dynamics is time translation invariant. This allows one to define energy, and by employing techniques of statistical physics,  temperature. These will be the basic ingredients together with the entropy associated with the amount of information.

\subsection{Force and inertia.}

Our starting assumption is directly motivated by Bekenstein's original thought experiment from which he obtained is famous entropy formula. He considered a particle with mass $m$ attached to a fictitious "string" that is lowered towards a black hole. Just before the horizon the particle is dropped in. Due to the infinite redshift the mass increase of the black hole can be made arbitrarily small, classically.   If one would take a thermal gas of particles, this fact would lead to problems with the second law of thermodynamics. Bekenstein solved this by arguing that when a particle is one Compton wavelength from the horizon, it is considered to be part of the black hole. Therefore, it increases the  mass and horizon area by a small amount, which he identified with one bit of information. This lead him to his area law for the black hole entropy. 

We want to mimic this reasoning not near a black hole horizon, but in flat non-relativistic space. So we consider a small piece of an holographic screen, and a particle of mass $m$  that approaches it from the side at which space time has already emerged. Eventually the particle  merge with the microscopic degrees of freedom on the screen, but before it does so, it already influences the amount of information that is stored on the screen. The situation is depicted in figure 2.

Motivated by Bekenstein's argument,  let us postulate that the change of entropy associated with the information on the boundary equals
\be
\label{postulate}
{\Delta S} = 2\pi k_B \qquad\quad{\mbox{when}} \qquad\quad
\Delta x= {\hbar \over mc}.
\ee
The reason for putting in the factor of $2\pi$, will become apparent soon. 
Let us rewrite this formula  in the slightly more general form by assuming that the change in entropy near the screen is linear in the displacement $\Delta x$. 
\be
\label{basiclaw}
\Delta S= 2\pi k_B  {mc\over\hbar} \Delta x.
\ee 
To understand why it is also proportional to the mass $m$, let us imagine splitting the particle into two or more lighter sub-particles. Each sub-particle then carries its own associated change in entropy after a shift $\Delta x$. Since entropy and mass are both additive, it is therefore natural that the entropy change is proportional to the mass. 
\begin{figure}[tbp]
\begin{center}
\includegraphics[scale=0.25]{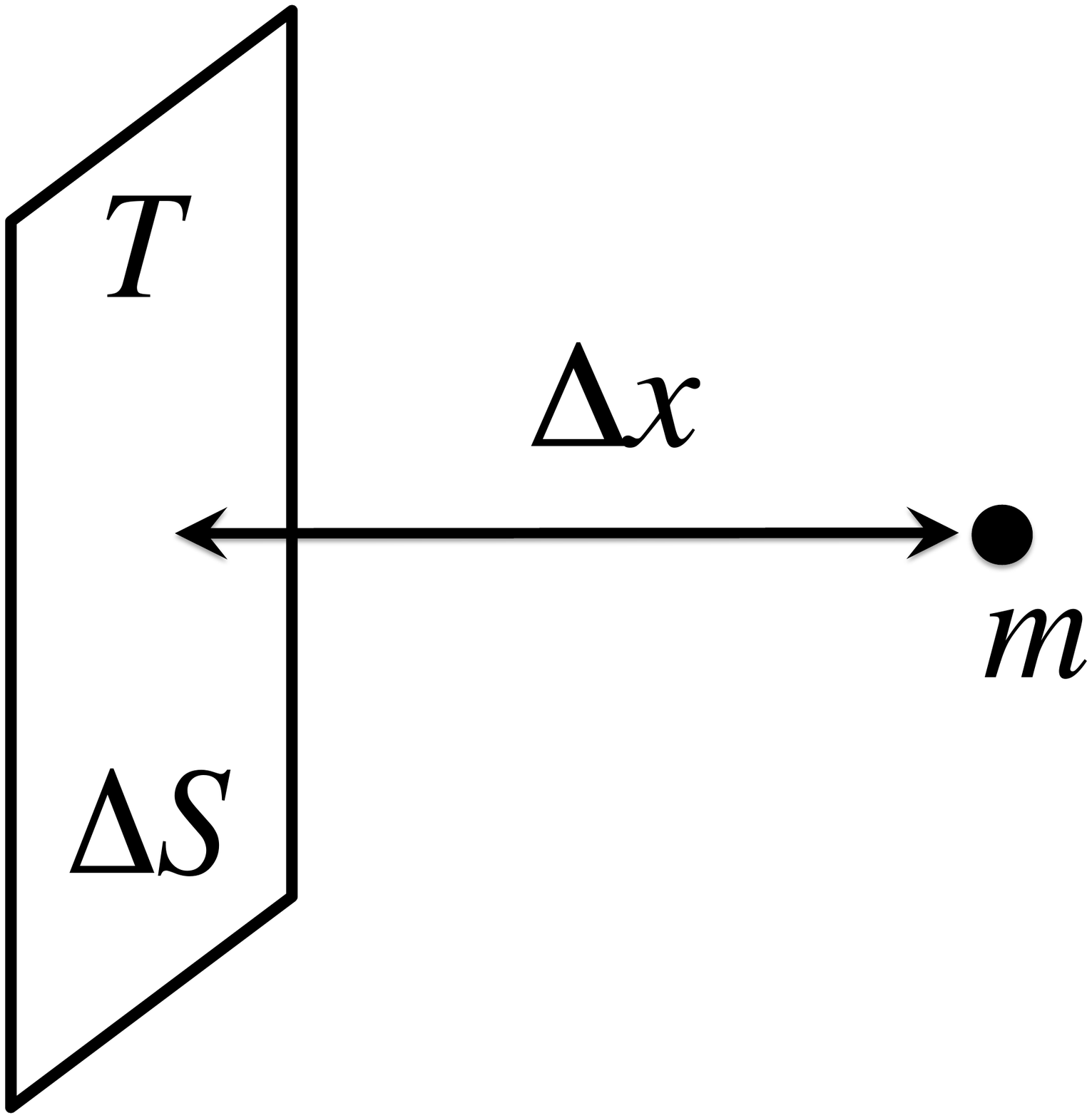}
\vspace{-0.5 cm}
\end{center}
\caption{\small{A particle with mass approaches a part of the holographic screen. The screen bounds the emerged part of space, which contains the particle, and stores data that describe the part of space that has not yet emerged, as well as some part of the emerged space.}}
\end{figure}  
How does force arise? The basic idea is to use the analogy with osmosis across a semi-permeable membrane.
When a particle has an entropic reason to be on one side of the membrane and the membrane carries a temperature, it will experience an effective force equal to
\be
\label{entropic}
 F \Delta x = T{\Delta S}.
\ee
This is the entropic force. Thus, in order to have a non zero force, we need to have a non vanishing temperature. From Newton's law we know that a force leads to a non zero acceleration.  Of course, it is well known that acceleration and temperature are closely related. Namely, as Unruh showed, an observer in an accelerated frame experiences a temperature
\be
\label{unruh}
 k_BT= {1\over 2\pi} {\hbar a\over  c},
\ee
where  $a$ denotes the acceleration. Let us take this as the temperature associated with the bits on the screen.
Now it is clear why the equation (\ref{basiclaw}) for $\Delta S$ was chosen to be of the given form, including the factor of $2\pi$. It is picked precisely in such a way that one recovers the second law of Newton
\be
\label{secondlaw}
F=ma.
\ee
as is easily verified by combining (\ref{unruh}) together with (\ref{basiclaw}) and (\ref{entropic}).  

Equation (\ref{unruh}) should be  read as a formula for the  temperature $T$  that is required to cause an acceleration equal to $a$.  And not as usual, as the temperature caused by an acceleration.

\bigskip

\subsection{Newton's law of gravity.}

Now suppose our boundary is not infinitely extended, but forms a closed surface. More specifically, let us assume it is a sphere. For the following it is best to forget about the Unruh law (\ref{unruh}), since we don't need it. It only served as a further motivation for (\ref{basiclaw}). The key statement is simply that we need to have a temperature in order to have a force. Since we want to understand the origin of the force,  we need to know where the temperature comes from.

One can think about the boundary as a storage device for information. Assuming that the holographic principle holds, the maximal storage space, or total number of bits, is proportional to the area $A$. In fact, in an theory of emergent space this how area may be defined: each fundamental bit occupies by definition one unit cell. 

Let us denote the number of used bits by $N$. It is natural to assume that this number will be proportional to the area. So we write
\be
\label{bits}
N  
={A c^3\over G\hbar}
\ee 
where we introduced a new constant $G$. Eventually this constant is going to be  identified with Newton's constant, of course.  But since we have not assumed anything yet about the existence a gravitational force, one can simply regard this equation as the definition of $G$.  So, the only assumption made here is that the number of bits is proportional to the area. Nothing more.

Suppose there is a total energy $E$ present in the system.  Let us now just make the simple assumption that the energy is divided evenly over the bits $N$.  The temperature is then determined by the equipartition rule
\be
\label{equipartition}
E = {1\over 2} N k_B T
\ee
as the average energy per bit.
 After this we need only one more equation:
\be
\label{E=Mc^2}
E=Mc^2.
\ee
Here $M$ represents the mass that would emerge in the part of space enclosed by the screen, see figure 3.  Even though the mass is not directly visible in the emerged space, its presence is noticed though its energy.

\begin{figure}[tbp]
\begin{center}
\includegraphics[scale=0.275]{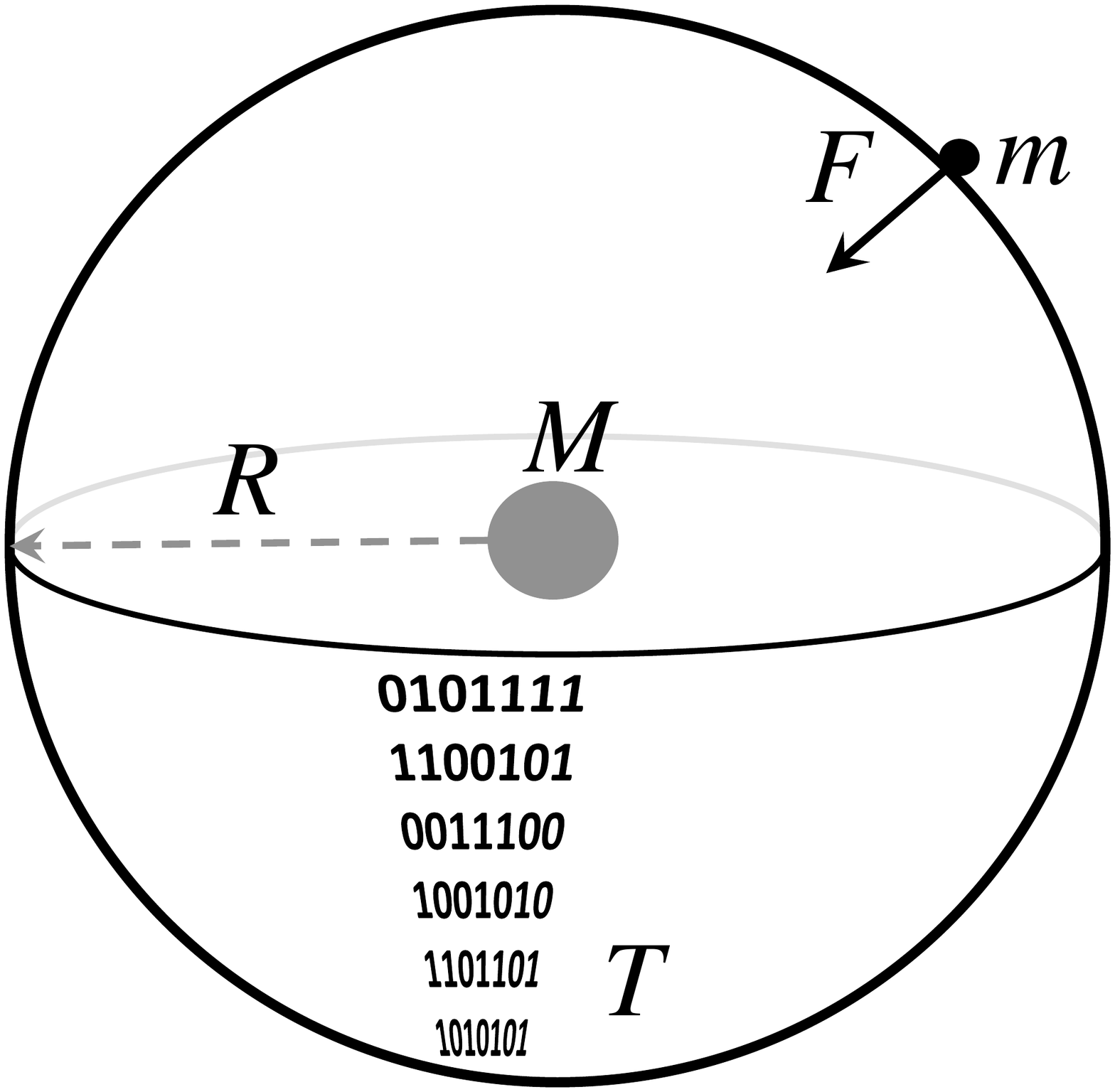}
\end{center}
\vspace{-0.4cm}
\caption{\small{A particle with mass $m$ near a spherical holographic screen. The energy is evenly distributed over the occupied bits, and is equivalent to the mass $M$ that would emerge in the part of space surrounded by the screen.} }
\end{figure}
The rest is straightforward: one eliminates $E$ and inserts the expression for the number of bits to determine $T$. Next one uses the postulate (\ref{basiclaw}) for the change of entropy to determine  the force. Finally one inserts  
$$A=4\pi R^2.$$ 
and one obtains the familiar law:
\be
\label{Newtonslaw}
F=G{Mm\over R^2}.
\ee
We have recovered Newton's law of gravitation, practically from first principles! 

These equations do not just come out by accident.  It had to work, partly for dimensional reasons, and also because the laws of Newton have been ingredients in the steps that lead to black hole thermodynamics and the holographic principle. In a sense we have reversed these arguments.   But the logic is clearly different, and sheds new light on the origin of gravity: it is an entropic force!  That is the main statement, which is new and has not been made before. If true, this should have profound consequences.

\bigskip

\subsection{Naturalness and robustness of the derivation.}

Our starting point was that space has one emergent holographic direction. The additional ingredients were that  (i) there is a change of entropy in the emergent direction (ii) the number of degrees of freedom are proportional to the area of the screen, and  (iii) the energy is evenly distributed over these degrees of freedom. After that it is unavoidable that the resulting force takes the form of Newton's law.  In fact, this reasoning can be generalized to arbitrary dimensions\footnote{In $d$ dimensions (\ref{bits}) includes a factor ${1\over 2}{ d\!-\!2\over d\!-\!3}$ to get the right identification with Newton's constant.} with the same conclusion.   But how robust and natural are these heuristic arguments? 

Perhaps the least obvious assumption is equipartition, which in general holds only for free systems.  But how essential is it?  Energy usually spreads over the microscopic degrees of freedom according to some non trivial distribution function. When the lost bits are randomly chosen among all bits,  one expects the energy change associated with $\Delta S$ still to be proportional to the energy per unit area $E/A$.  This fact could therefore be true even when equipartition is not strictly obeyed. 

Why do we need the speed of light $c$ in this non relativistic context? It was necessary to translate the mass $M$ in to an energy, which provides the heat bath required for the entropic force.  In the non-relativistic setting this heat bath is infinite, but in principle one has take into account that the heat bath loses or gains energy when the particle changes its location under the influence of an external force. This will lead to relativistic redshifts, as we will see.   

Since the postulate (\ref{postulate}) is the basic assumption from which everything else follows, let us discuss its meaning in more detail. Why does the entropy precisely change like this when one shifts by one Compton wave length? In fact, one may wonder why we needed to introduce Planck's constant in the first place, since the only aim was to derive the classical laws of Newton. Indeed,  $\hbar$ eventually drops out of the most important formulas.  So, in principle one could multiply it with any constant and still obtain the same result. 
Hence,  $\hbar$ just serves as an {\it auxiliary} variable that is needed for dimensional reasons. It can therefore be chosen at will, and defined so that (\ref{postulate}) is exactly valid.  The main content of this equation is therefore simply that there is an entropy change perpendicular to the screen proportional to the mass $m$ and the displacement $\Delta x$.  That is all there is to it.

If we would move further away from the screen, the change in entropy will in general no longer be given by the same rule. Suppose the particle stays at radius $R$ while the screen is moved to $R_0<R$. The number of bits on the screen is multiplied by a factor  $(R_0/R)^2$, while the temperature is divided by the same factor. Effectively, this means that only $\hbar$ is multiplied by that factor, and since it drops out,  the resulting force will stay the same. In this situation the information associated with the particle is no longer concentrated in a small area, but spreads over the screen. The next section contains a proposal for precisely how it is distributed, even for general matter configurations.

\subsection{Inertia and the Newton potential.}  

To complete the derivation of the laws of Newton we have to understand why the symbol $a$, that was basically introduced by hand in (\ref{unruh}), is equal to the physical acceleration $\ddot{x}$. In fact, so far our discussion was quasi static, so we have not determined yet how to connect space at different times. 
In fact,  it may appear somewhat counter-intuitive that the temperature $T$ is related to the vector quantity $a$, while in our identifications the entropy gradient $\Delta S/\Delta x$ is related to the scalar quantity $m$. In a certain way it seems more natural to have it the other way around. 

So let reconsider what happens to the particle with mass $m$ when it approaches the screen. Here it should merge with the microscopic degrees of freedom on the screen, and hence it will be made up out of the same bits as those that live on the screen. Since each bit carries an energy ${1\over 2}k_BT$, the number of bits $n$ follows from 
\be
mc^2= {1\over 2}n\, k_BT.
\ee   
When we insert this in to equation (\ref{basiclaw}), and use (\ref{unruh}), we can express the entropy change in terms of the acceleration as
\be
\label{gradient}
{\Delta S \over  n} \,  = k_B\,{a \,\Delta x \over 2c^2} 
\ee
By combining the above equations one of course again recovers  $F=ma$ as the entropic force. But,  by introducing the number of bits $n$ associated with the particle, we succeeded in making the identifications more natural in terms of their scalar versus vector character. In fact, we have eliminated $\hbar$ from the equations, which in view of our earlier comment is a good thing. 

Thus we conclude that acceleration is related to an entropy gradient. This will be one of our main principles:  inertia is a consequence of the fact that a particle in rest will stay in rest because there are no entropy gradients.  
Given this fact it is natural to introduce the Newton potential $\Phi$ and write the acceleration as a gradient 
$$
a =-\nabla \Phi.
$$
This allows us to express the change in entropy in the concise way
\be
\label{deltaS}
{\Delta S\over n} = -\, k_B  \,{ \Delta \Phi\over 2 c^2}
\ee
We thus reach the important conclusion that the Newton potential $\Phi$ keeps track of the depletion of the entropy per bit. It is therefore natural to identify it with a coarse graining variable, like the (renormalization group) scale in AdS/CFT. Indeed, in the next section we propose a holographic scenario for the emergence of space in which the Newton potential precisely plays that role.  This allows us to generalize our discussion to other mass distributions and arbitrary positions in a natural and rather beautiful way, and give additional support  for the presented arguments.

\section{Emergent Gravity for General Matter Distributions.}

Space emerges at a macroscopic level only after coarse graining. Hence,  there will be a finite entropy associated with each matter configuration.  This entropy measures the amount of microscopic information that is invisible to the macroscopic observer. In general, this amount will depend on the distribution of the matter. The information is being processed by the microscopic dynamics, which looks random from a macroscopic point of view. But to determine the force we don't need the details of the information, nor the exact dynamics, only the amount of information given by the entropy, and the energy that is associated with it. If the entropy changes as a function of the location of the matter distribution,  it will lead to an entropic force. 

Therefore, space can not just emerge by itself. It has to be endowed by a book keeping device that keeps track of the amount of information for a given energy distribution. It turns out, that in a non relativistic situation this device is provided by Newton's potential $\Phi$.  And the resulting entropic force is called gravity. 

We start from microscopic information. It is assumed to be stored on holographic screens.   Note that information has a natural inclusion property: by forgetting certain bits, by coarse graining, one reduces the amount of information. This coarse graining can be achieved through averaging, a block spin transformation, integrating out, or some other renormalization group procedure. At each step one obtains a further coarse grained version of the original microscopic data. 
The gravitational or closed string side of these dualities is by many still believed to be independently defined. But in our view these are macroscopic theories, which by chance we already knew about before we understood they were the dual of a microscopic theory without gravity. We can't resist making the analogy with a situation in which we would have developed a theory for elasticity using stress tensors in a continuous medium half a century before knowing about atoms. We probably would have been equally resistant in accepting the obvious.  Gravity and closed strings are not much different, but we just have not yet got used to the idea. 

The coarse grained data live on smaller screens obtained by moving the first screen further into the interior of the space. The information that is removed by coarse graining is replaced by the emerged part of space between the two screens.    In this way one gets a nested or foliated description of space by having surfaces contained within surfaces. In other words, just like in AdS/CFT, there is one emerging direction in space that corresponds to a "coarse graining" variable, something like the cut-off scale of the system on the screens. 

A priori there is no preferred holographic direction in flat space.   However, this is where we use our observation about the Newton potential. It is the natural variable that measures the amount of coarse graining on the screens. Therefore, the holographic direction is given by the gradient $\nabla\Phi$ of the Newton potential.   In other words, the holographic screens correspond to equipotential surfaces. This leads to a well defined foliation of space, except that screens may break up in to disconnected parts that each enclose different regions of space.   This is depicted in figure 4.

The amount of coarse graining is measured by the ratio $-\Phi/2c^2$, as can be seen from (\ref{deltaS}).  Note that this is a dimensionless number that is always between zero and one. It is only equal to one on the horizon of a black hole. We interpret this as the point where all bits have been maximally coarse grained.  Thus the foliation naturally stops at black hole horizons.

\begin{figure}[tbp]
\begin{center}
\includegraphics[scale=0.25]{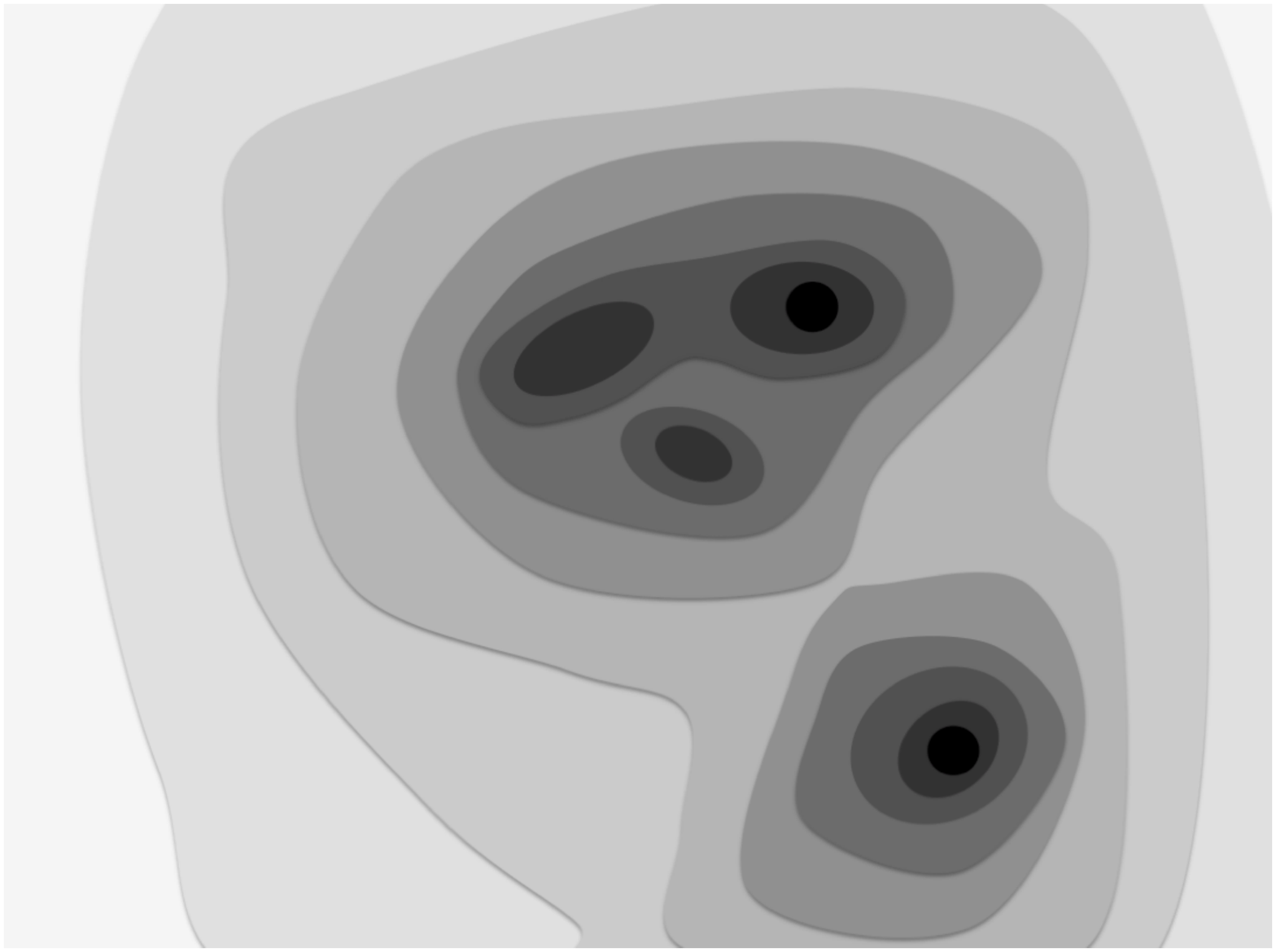}
\vspace{-0.0cm}
\caption{\small{The holographic screens are located at equipotential surfaces. The information on the screens is coarse grained in the direction of decreasing values of  the Newton potential $\Phi$. The maximum coarse graining happens at black hole horizons, when $\Phi/2c^2=-1$.}}
\vspace{-0.4cm}
\end{center}
\end{figure}

\subsection{The Poisson equation for general matter distributions.}  

Consider a microscopic state, which after coarse graining corresponds to a given mass distribution in space. All microscopic states that lead to the same mass distribution belong to the same macroscopic state. The entropy for each of these state is defined as the number of microscopic states that flow to the same macroscopic state.  

We want to determine the gravitational force  by using virtual displacements, and calculating the associated change in energy. So, let us freeze time and keep all the matter at fixed locations.  Hence,  it is described by a static matter density $\rho(\vec{r})$. Our aim is to obtain the force that the matter distribution exerts on a collection of test particles with masses $m_i$ and positions $\vec{r}_i$.

We choose a holographic screen $\cal S$ corresponding to an equipotential surface with fixed Newton potential $\Phi_0$. We assume that the entire mass distribution given by $\rho(x)$ is contained inside the volume enclosed by the screen, and all test particles are outside this volume. To explain the force on the particles, we again need to determine the work that is performed by the force and show that it is naturally written as the change in entropy multiplied by the temperature. The difference with the spherically symmetric case is that the temperature on the screen is not necessarily constant. Indeed, the situation is in general not in equilibrium. Nevertheless, one can locally define temperature and entropy per unit area.

First let us identify the temperature. We do this by taking a test particle and moving it close to the screen, and measuring the local acceleration. Thus, motivated by our earlier discussion we define temperature analogous to (\ref{unruh}), namely by
\be
k_BT={1\over 2\pi}{\hbar \nabla \Phi\over k c}. 
\ee
Here the derivative is taken in the direction of the outward pointing normal to the screen.   
Note at this point $\Phi$ is just introduced as a device to describe the local acceleration, but we don't know yet whether it satisfies an equation that relates it to the mass distribution.
 
The next ingredient is the density of bits on the screen. We again assume that these bits are uniformly distributed, and so (\ref{bits}) is generalized to
\be
\label{bitdensity}
dN={c^3\over G\hbar}\, dA.
\ee
Now let us impose the analogue of the equipartition relation (\ref{equipartition}). It is takes the form of an integral expression for the energy 
\be 
E= {1\over 2} k_B  \int_{\cal S}  T dN.
\ee
It is an amusing exercise to work out the consequence of this relation.  
Of course, the energy $E$ is again expressed in terms of the total enclosed mass $M$. 
 After inserting our identifications for the left hand side one obtains a familiar relation:  Gauss's law! 
\be
M={1\over 4\pi G} \int_{\cal S}\! \nabla\Phi\cdot dA.
\ee
This should hold for arbitrary screens given by equipotential surfaces. When a bit of mass is added to the region enclosed by the screen $\cal S$, for example, by first putting it close to the screen and then pushing it across, the mass $M$ should change accordingly. This condition can only hold in general if the potential $\Phi$ satisfies
 the Poisson equation
\be
 \nabla^2\Phi(\vec{r})=  4\pi G\,\rho(\vec{r}).
\ee
We conclude that by making natural identifications for the temperature and the information density on the holographic screens, that the laws of gravity come out in a straightforward fashion.

\bigskip

\subsection{The gravitational force for arbitrary particle locations.}

The next issue is to obtain the force acting on matter particles that are located at arbitrary points outside the screen.
For this we need a generalization of the first postulate (\ref{basiclaw}) to this situation. What is the entropy change due to arbitrary infinitesimal displacements $\delta\vec{r}_i$ of the particles? There is only one natural choice here.  We want to find the change $\delta s$ in the entropy density 
locally on the screen $\cal S$.   We noted in (\ref{deltaS}) that the Newton potential $\Phi$ keeps track of the changes of information per unit bit. Hence, the right identification for the change of entropy density is
\be
\, \delta\!{}_{\, }s \, = k_B {\delta \Phi\over 2c^2} \, d N
\ee
where $\delta\Phi$ is the response of the Newton potential due to the shifts $\delta \vec{r}_i$ of the positions of the particles. To be specific, $\delta\Phi$ is determined by solving the variation of the Poisson equation
\be
\nabla^2\delta\Phi(\vec{r})=4\pi G\sum_i m_i\, \delta \vec{r}_i\!\cdot\!\nabla_i\, \delta(\vec{r}-\vec{r}_i)
\ee
One can verify that with this identification one indeed reproduces  the entropy shift (\ref{basiclaw}) when one of the  particles approaches the screen. 

\begin{figure}[tbp]
\begin{center}
\includegraphics[scale=0.30]{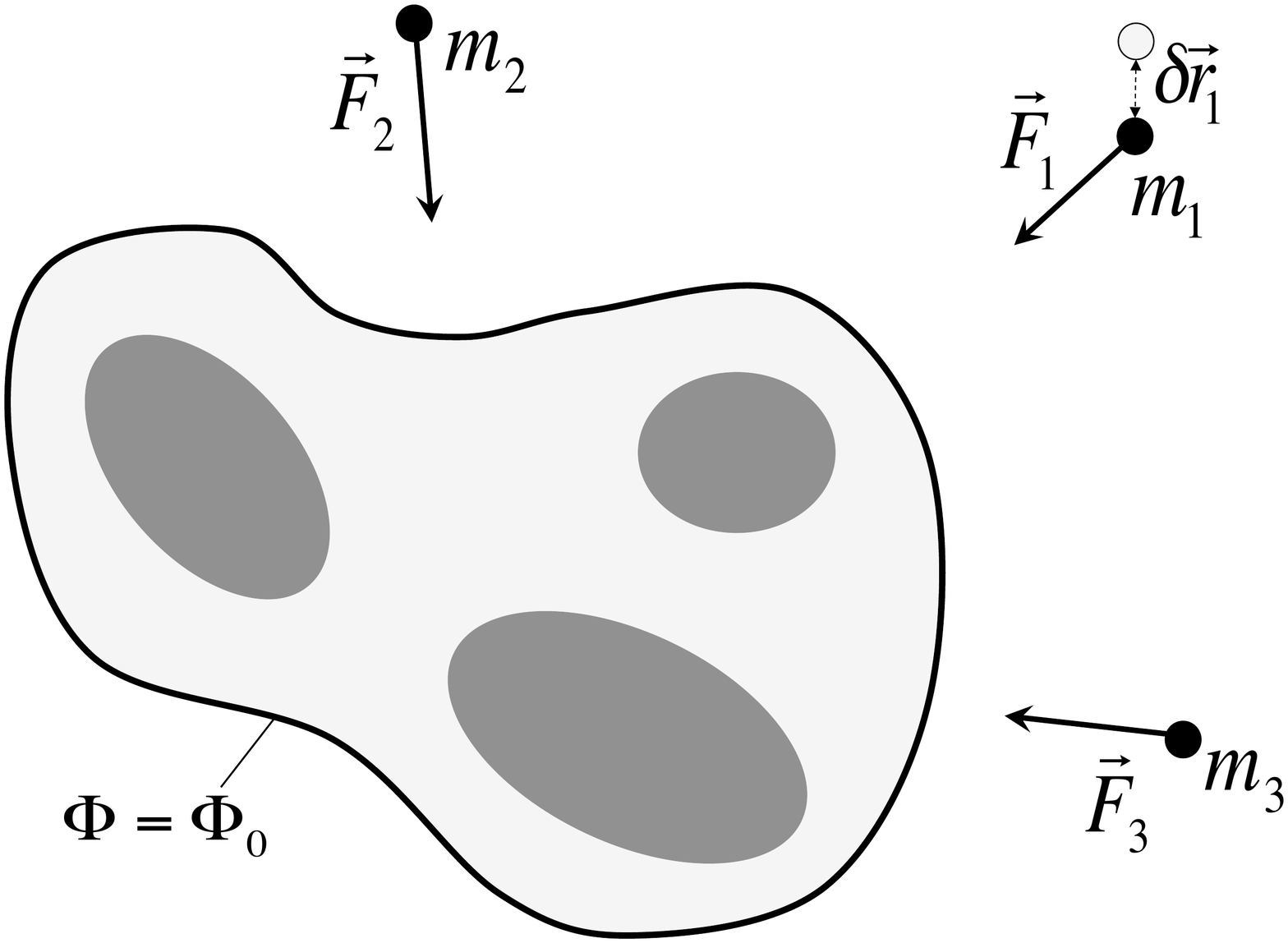}
\vspace{-0.4 cm}
\end{center}
\caption{\small{A general mass distribution inside the not yet emerged part of space enclosed by the screen. A collection of test  particles with masses $m_i$ are located at arbitrary points $\vec{r}_i$ in the already emerged space outside the screen. The forces $\vec{F}_i$ due to gravity are determined by the virtual work done after infinitesimal displacement $\delta\vec{r}_i$ of the particles. }}
\end{figure}  

Let us now determine the entropic forces on the particles. The combined work done by all of the forces on the test particles is determined by the first law of thermodynamics. However, we need to express in terms of the local temperature and entropy variation. Hence,
\be
\sum_i \,\vec{F}_i\cdot\delta\vec{r}_i = \int_{\cal S}T\, \delta s 
\ee
To see that this indeed gives the gravitational force in the most general case, one simply has to use the electrostatic analogy. Namely,  one can redistribute the entire mass $M$ as a mass surface density over the screen $\cal S$ without changing the forces on the particles. The variation of the Newton potential can be obtained from the Greens function for the Laplacian. The rest of the proof is a straightforward application of electrostatics, but then applied to gravity.   The basic identity one needs to prove is
\be
\sum_i \,\vec{F}_i\cdot\delta \vec{r}_i = {1\over 4\pi G}\int_{\cal S} \bigl(\delta \Phi \nabla\Phi-\Phi \nabla\delta\Phi\bigr)\,dA
\ee
which holds for any location of the screen outside the mass distribution. This is easily verified by using Stokes theorem and the Laplace equation.
The second term vanishes when the screen is chosen at a equipotential surface. To see this, simply replace $\Phi$ by  $\Phi_0$ and pull it out of the integral. Since $\delta\Phi$  is sourced by only the particles outside the screen,  the remaining integral just gives zero. 

The forces we obtained are independent of the choice of the location of the screen. We could have chosen any equipotential surface, and we would obtain the same values for $\vec{F}_i$, the ones described by the laws of Newton.  That all of this works is not just a matter of dimensional analysis. 
The invariance under the choice of equipotential surface is very much consistent with idea that a particular location corresponds to an arbitrary choice of the scale that controls the coarse graining of the microscopic data.  The macroscopic physics, in particular the forces, should be independent of that choice.

\section{ The Equivalence Principle and the Einstein equations.}

Since we made use of the speed of light $c$ in our arguments, it is a logical step to try and generalize our discussion to a relativistic situation. 
 So let us assume that the microscopic theory knows about Lorentz symmetry, or even has the Poincar\'{e} group as a global symmetry.  This means we have to combine time and space in to one geometry.
A scenario with emergent space-time quite naturally leads to general coordinate invariance and curved geometries,  since a priori there are no preferred choices of coordinates, nor a reason why curvatures would not be present.  Specifically, we would like to see how Einstein's general relativity emerges from similar reasonings as in the previous section. We will indeed show that this is possible. But first we study the origin of inertia and the equivalence principle.

 \subsection{The law of inertia and the equivalence principle.}

Consider a static background with a global time like Killing vector $\xi^a$. To see the emergence of inertia and the equivalence principle, one has to relate the choice of this Killing vector field  with the temperature and the entropy gradients. In particular, we like to see that the usual geodesic motion of particles can be understood as being the result of an entropic force.

In general relativity\footnote{In this subsection and the next we essentially follow Wald's book on general relativity (pg.\ 288-290).  We use a notation in which $c$ and $k_B$ are put equal to one, but we will keep $G$ and $\hbar$ explicit.} the natural generalization of Newton's potential is \cite{Waldbook}, 
\be
\label{phi}
\phi ={1\over 2} \log(-\xi^a\xi_a).
\ee  
Its exponent $e^\phi$ represents the redshift factor that relates the local time coordinate to that at a reference point with $\phi=0$, which we will take to be at infinity.  

Just like in the non relativistic case, we like to use $\phi$ to define a foliation of space, and put our holographic screens at  surfaces of constant redshift. This is a natural choice, since in this case the entire screen uses the same time coordinate. So the processing of the microscopic data on the screen can be done using signals that travel without time delay. 

We want to show that the redshift  perpendicular to the screen can be understood microscopically  as originating from the entropy gradients\footnote{In this entire section it will be very useful to keep the polymer example of section 2 in mind, since that will make the logic of our reasoning very clear.} .
To make this explicit, let us consider the force that acts on a particle of mass $m$. In a general relativistic setting force is less clearly defined, since it can be transformed away by a general coordinate transformation. But by using the time-like Killing vector one can given an invariant meaning to the concept of force \cite{Waldbook}.   

The four velocity $u^a$ of the particle and its acceleration  $a^b\!\equiv\!u^a\nabla_a u^b$  can be expressed in terms of the Killing vector $\xi^b$ as
$$
u^b =e^{-\phi}\xi^b, \qquad\qquad a^b=e^{-2\phi}\xi^a\nabla_a\xi^b.
$$
We can further rewrite the last equation by making use of the Killing equation 
$$
\nabla_a\xi_b+\nabla_b\xi_a=0
$$ 
and the definition of $\phi$.  One finds that the acceleration can again be simply expressed the gradient
\be
\label{relaccel}
a^b  =-\nabla^b\phi.
\ee
Note that just like in the non relativistic situation the acceleration is perpendicular to screen $\cal S$. So we can turn it in to a scalar quantity by
contracting it with a unit outward pointing vector $N^b$ normal to the screen $\cal S$ and to $\xi^b$. 

The local temperature $T$ on the screen is now in analogy with the non relativistic situation defined by
\be
\label{Trel}
T = {\hbar \over 2\pi} e^{\phi} N^b\nabla_b\phi.
\ee
Here we inserted a redshift factor $e^\phi$, because the temperature $T$  is measured with respect to the reference point at infinity. 

To find the force on a particle that is located very close to the screen, we first use again the same postulate as in section two. Namely, we assume that the change of entropy at the screen is $2\pi$ for a displacement by one Compton wavelength normal to the screen. Hence,
 \be
 \nabla_a S = -2\pi {m\over \hbar} N_a, 
 \ee
 where the minus sign comes from the fact that the entropy increases when we cross from the outside to the inside. The comments made on the validity of this postulate in section 3.3 apply here as well. 
 The entropic force now follows from (\ref{Trel}) 
 \be
 \label{entropicforce}
 F_a =T\nabla_a S = -m e^\phi\nabla_a\phi
 \ee
 This is indeed the correct gravitational force that is required to keep a particle at fixed position near the screen, as measured from the reference point at infinity.  
It is the relativistic analogue of Newton's law of inertia $F=ma$. The additional factor $e^\phi$ is due to the redshift. Note that $\hbar$ has again dropped out. 
 
It is instructive to rewrite the force equation (\ref{entropicforce}) in a microcanonical form. Let $S(E,x^a)$ be the total entropy  associated with a system with total energy $E$ that contains a particle with mass $m$ at position $x^a$. Here $E$ also includes the energy of the particle.  The entropy will in general also dependent on many other parameters, but we suppress these in this discussion.

As we explained in the section 2, an entropic force can be determined micro-canonically by adding by hand an external force term, and impose that the entropy is extremal.  For this situation this condition looks like 
\be
\label{extremal}
{d\over dx^a} \,S\bigl(E\!+\!e^{\phi(x)}m,\,x^a\bigr)=0.
\ee
One easily verifies that this leads to the same equation (\ref{entropicforce}). This fixes the equilibrium point where the {\it external} force, parametrized by $\phi(x)$ and the {\it entropic} force statistically balance each other. Again we stress the point that there is no {\it microscopic} force acting here!  The analogy with equation (\ref{extremal1}) for the polymer, discussed in section 2,  should be obvious now.

Equation (\ref{extremal}) tells us that the entropy remains constant if  we move the particle and simultaneously reduce its energy  by the redshift factor. This is true only when the particle is very light, and 
does not disturb the other energy distributions. It simply serves as a probe of the emergent geometry.  This also means that redshift function $\phi(x)$ is  entirely fixed by the other matter in the system.

We have arrived at equation (\ref{extremal})  by making use of the identifications of the temperature and entropy variations in space time. But actually we should have gone the other way. We should have started from the microscopics and defined the space dependent concepts in terms of them. We chose not to follow such a presentation, since it might have appeared somewhat contrived. 

But it is important to realize that the redshift must be seen as a consequence of the entropy gradient and not the other way around. The equivalence principle tells us that redshifts can be interpreted in the emergent space time as either due to a gravitational field or due to the fact that one considers an accelerated frame. Both views are equivalent in the relativistic setting, but neither view is microscopic. Acceleration and gravity are both emergent phenomena.

\subsection{Derivation of the Einstein equations.}

We now like to extend our derivation of the laws of gravity to the relativistic case, and obtain the Einstein equations. This can indeed be done in natural and very analogous fashion. So we again consider a holographic screen on closed surface of constant redshift $\phi$. We assume that it is enclosing a certain static mass configuration with total mass $M$.
The bit density on the screen is again given by
\be
dN= {dA\over G\hbar}
\ee 
as in (\ref{bitdensity}). Following the same logic as before,  let us assume that the energy associated with the mass $M$  is distributed over all the bits.  Again by equipartition each bit carries a mass unit equal to ${1\over 2}T$.
Hence
\be 
M={1\over 2} \int_{\cal S}  T dN
\ee
After inserting the identifications for $T$ and $dN$ we obtain 
\be
\label{Komar1}
M ={1\over 4\pi G}\int_{\cal S}  e^{\phi} \nabla \phi\cdot dA
\ee 
Note that again $\hbar$ drops out, as expected.
The equation (\ref{Komar1}) is indeed known to be the natural generalization of Gauss's law to General Relativity.  Namely, the right hand side is precisely Komar's definition of the mass contained inside an arbitrary volume inside any static curved space time. It can be derived by assuming the Einstein equations. 
In our reasoning, however, we are coming from the other side. We made identifications for the temperature and the number of bits on the screen. But we don't know yet whether it satisfies any field equations. The key question at this point is whether the equation (\ref{Komar1}) is sufficient to derive the full Einstein equations. 

An analogous question was addressed by Jacobson for the case of null screens. By adapting his reasoning to this situation, and combining it with Wald's exposition of the Komar mass,  it is straightforward to construct an argument that naturally leads to the Einstein equations. We will present a sketch of this.

First we make use of the fact that the Komar mass can be re-expressed in terms of the Killing vector $\xi^a$ as
\be
M ={1\over 8\pi G}\int_{\cal S}   dx^a \!\wedge \!dx^b\,  \epsilon_{abcd}  \nabla^{c}\xi^{d}
\ee 
 The left hand side can be expressed as an integral over the enclosed volume of certain components of stress energy tensor $T_{ab}$.  For the right hand side one first uses Stokes theorem and subsequently the relation
\be
\nabla^a\nabla_a\xi^b=-R^{b}{}_a\xi^a
\ee
which is implied by the Killing equation for $\xi^a$. 
This leads to the integral relation \cite{Waldbook}
\be
\label{Einstein}
2\int_{\Sigma}\left(T_{ab} -{1\over 2}Tg_{ab}\right) n^a \xi^b dV= {1\over 4\pi G}\int_{\Sigma}R_{ab} n^a \xi^b dV
\ee
where $\Sigma$ is the three dimensional volume bounded by the holographic screen $\cal S$ 
and $n^a$ is its normal.   The particular combination of the stress energy tensor on the left hand side can presumably be fixed by comparing properties on both sides, such as for instance the conservation laws of the tensors that occur in the integrals.

This equation is derived in a general static background with a time like Killing vector $\xi^a$. By requiring that it holds for arbitrary screens would imply that also the integrands on both sides are equal.  This gives us only a certain component of the Einstein equation. In fact, we can choose the surface $\Sigma$ in many ways, as long as its boundary is given by $\cal S$. This means that we can vary the normal $n^a$. But that still leaves a contraction with the Killing vector $\xi^a$. 

To get to the full Einstein we now use a similar reasoning as Jacobson \cite{Jacobson}, except now applied to time-like screens.   Let us consider a very small region of space time and look also at very short time scales. Since locally every geometry looks approximately like Minkowski space, we can choose approximate time like Killing vectors. Now consider a small local part of the screen, and impose that when matter crosses it, the value of the Komar integral will jump precisely by the corresponding mass $m$.  By following the steps described above then leads to (\ref{Einstein}) for all these Killing vectors, and for arbitrary screens. This is sufficient to obtain the full Einstein equations.

\subsection{The force on a collection of particles at arbitrary locations.}

We close this section with explaining the way the entropic force acts on a collection of particles  at arbitrary locations $x_i$ away from the screen and the mass distribution. The Komar definition of the mass will again be useful for this purpose.  The definition of the Komar mass depends on the choice of Killing vector $\xi_a$. In particular also in its norm, the redshift factor $e^\phi$. If one moves the particles by a virtual displacement $\delta x_i$, this will effect the definition of the Komar mass of the matter configuration inside the screen. In fact, the temperature on the screen will be directly affected by a change in the redshift factor.  

The virtual displacements can be performed quasi statically, which means that the Killing vector itself is still present. Its norm, or the redshift factor, may change, however. In fact, also the spatial metric may be affected by this displacement. We are not going to try to solve these dependences, since that would be impossible due to the non linearity of the Einstein equations. But we can simply use the fact that the Komar mass is going to be a function of the positions $x_i$ of the particles.

Next let us assume that in addition to this $x_i$ dependence of the Komar mass that the entropy of the entire system has also explicit $x_i$ dependences, simply due to changes in the amount of information. These are de difference that will lead to the entropic force that we wish to determine. We will now give a natural prescription for the entropy dependence that is based on a maximal entropy principle and indeed gives the right forces.
Namely, assume that the entropy may be written as a function of the Komar mass $M$ and in addition on the $x_i$. But since the Komar mass  should be regarded as a function $M(x_i)$ of the positions $x_i$, there will be an explicit and an implicit $x_i$ dependence in the entropy. The maximal entropy principle implies that these two dependences should cancel. So we impose
\be
\label{extreem}
S\bigl(M(x_i\!+\!\delta x_i) ,\ x_i\!+\!\delta x_i\bigr)= S\bigl(M(x_i), x_i\bigr)
\ee
By working out this condition and singling out the equations implied by each variation $\delta x^i$ one finds 
\be
\nabla_iM+T \nabla_i S=0
\ee
The point is that the first term represents the force that acts on the $i$-th particle due to the mass distribution inside the screen. This force is indeed equal to minus the derivative of the Komar mass, simply because of energy conservation. But again, this is not the microscopic reason for the force. In analogy with the polymer, the Komar mass represents the energy of the heat bath.  Its dependence on the position of the other particles is caused by redshifts whose microscopic origin lies in the depletion of the energy in the heat bath due to entropic effects.  

Since the Komar integral is defined on the holographic screen, it is clear that like in the non relativistic case  the force is expressed as an integral over this screen as well.  We have not tried to make this representation more concrete. Finally, we note that this argument was very general, and did not really use the precise form of the Komar integral, or the Einstein equations. So it should be straightforward to generalize this reasoning to higher derivative gravity theories by making use of Wald's Noether charge formalism \cite{WaldNoether}, which is perfectly suitable for this purpose.

\section{Conclusion and Discussion}
The ideas and results presented in this paper lead to many questions. In this section we discuss and attempt to answer some of these. First we present our main conclusion. 

\subsection{The end of gravity as a fundamental force.}

Gravity has given many hints of being an emergent phenomenon, yet up to this day it is still seen as a fundamental force. The similarities with other known emergent phenomena, such as thermodynamics and hydrodynamics, have been mostly regarded as just suggestive analogies.   It is time we not only notice the analogy, and talk about the similarity, but finally do away with gravity as a fundamental force. 

Of course, Einstein's geometric description of gravity is beautiful, and in a certain way compelling. Geometry appeals to the visual part of our minds, and is amazingly powerful in summarizing many aspects of a physical problem.  Presumably this explains why we, as a community, have been so reluctant to give up the geometric formulation of gravity as being fundamental. But it is inevitable we do so. If gravity is emergent, so is space time geometry. Einstein tied these two concepts together, and both have to be given up if we want to understand one or the other at a more fundamental level. 

The results of this paper suggest gravity arises as an entropic force, once space and time themselves have emerged.  If the gravity and space time can indeed be explained as emergent phenomena, this should have important implications for many areas in which gravity plays a central role. It would be especially interesting to investigate the consequences for cosmology. For instance, the way redshifts arise from entropy gradients could lead to many new insights. 

The derivation of the Einstein equations presented in this paper is analogous to previous works, in particular \cite{Jacobson}. 
Also other authors have proposed that gravity has an entropic or thermodynamic origin, see for instance  \cite{Padma}. 
But we have added an important element that is new. Instead of only focussing on the equations that govern the gravitational field,  we uncovered what is the origin of force and inertia in a context in which space is emerging. We identified a cause, a mechanism, for gravity. It is  driven by differences in entropy, in whatever way defined,  and a consequence of the statistical averaged random dynamics at the microscopic level. The reason why gravity has to keep track of energies as well as entropy differences is now clear. It has to, because this is what causes motion!

The presented arguments have admittedly been rather heuristic. One can not expect otherwise, given the fact that we are entering an unknown territory in which space does not exist to begin with. The profound nature of these questions in our view justifies the heuristic level of reasoning.  The assumptions we made have been natural: they fit with existing ideas  and are supported by several pieces of evidence. 
In the following we gather more supporting evidence from string theory, the AdS/CFT correspondence, and black hole physics.

\subsection{Implications for string theory and relation with AdS/CFT.}

If gravity is just an entropic force, then what does this say about string theory? Gravity is seen as an integral part of string theory, that can not be taken out just like that.  
But we do know about dualities between closed string theories that contain gravity and decoupled open string theories that don't.  A particularly important example  is the AdS/CFT correspondence. 

The open/closed string and AdS/CFT correspondences are manifestations of the UV/IR connection that is deeply engrained within string theory. This connection implies that short and long distance physics can not be seen as totally decoupled. Gravity is a long distance phenomenon that clearly knows about short distance physics, since it is evident that Newton's constant is a measure for the number of microscopic degrees of freedom.  String theory invalidates the "general wisdom" underlying the Wilsonian effective field theory,  namely that integrating out short distance degrees of freedom only generate local terms in the effective action, most of which become irrelevant at low energies. If that were completely true, the macroscopic physics would be insensitive to the short distance physics. 

The reason why the Wilsonian argument fails, is that it makes a too conservative assumption about the asymptotic growth of the number of states at high energies. In string theory the number of high energy open string states is such that integrating them out indeed leads to long range effects. Their one loop amplitudes are equivalent to the tree level contributions due to the exchange of close string states, which among other are responsible for gravity. This interaction is, however, equivalently represented by the sum over all quantum contributions of the open string. In this sense the emergent nature of gravity is also supported by string theory.

The AdS/CFT correspondence has an increasing number of applications to areas of physics in which gravity is not present at a fundamental level. Gravitational techniques are used as tools to calculate physical quantities in a regime where the microscopic description fails.  The latest of these developments is the application to condensed matter theory. No one doubts that in these situations gravity emerges only as an effective description. It arises not in the same space as the microscopic theory, but in a holographic scenario with one extra dimension.  No clear explanation exists of where this gravitational force comes from. The entropic mechanism described in this paper should be applicable to these physical systems, and explain the emergence of gravity.

\begin{figure}[tbp]
\begin{center}
\includegraphics[scale=0.275]{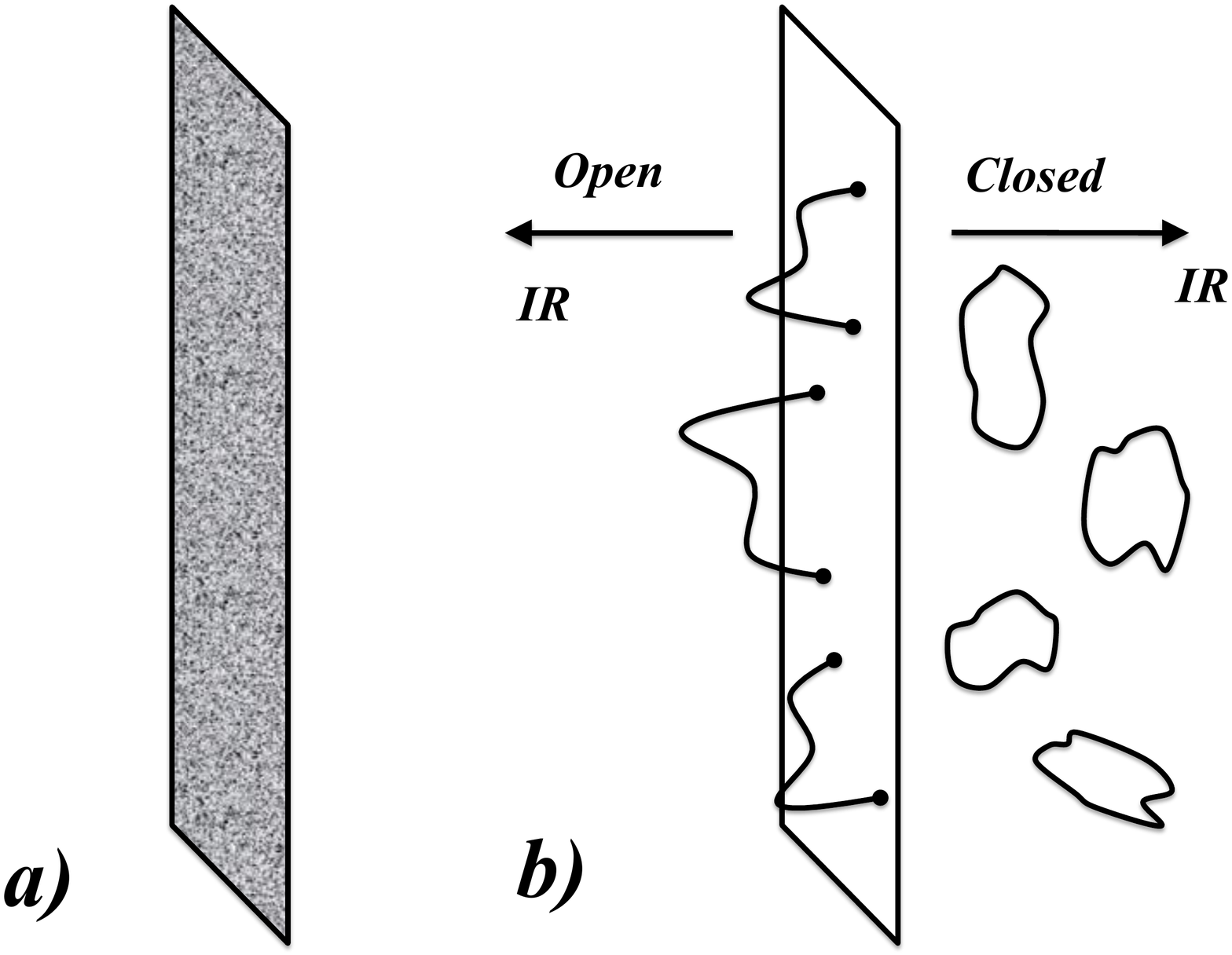}
\vspace{-0.2cm}
\caption{\small{The microscopic theory in a) is effectively described by a string theory consisting of open and closed strings  as shown in b). Both types of strings are cut off in the UV.
 }}
 \vspace{-0.6cm}
\end{center}
\end{figure}

The holographic scenario discussed in this paper has certainly be inspired by the way holography works in AdS/CFT and open closed string correspondences. In string language, the holographic screens can be identified with D-branes, and the microscopic degrees of freedom on these screens represented as open strings defined with a cut off in the UV. The emerged part of space is occupied by closed strings, which are also defined with a UV cut off, as shown in figure 6. The open and closed string cut offs are related by the UV/IR correspondence: pushing the open string cut off to the UV forces the closed string cut off towards the IR, and vice versa.  The value of the cut offs is determined by the location of the screen. Integrating out the open strings produces the closed strings, and leads to the emergence of space and gravity.  Note, however, that from our point of view the existence of gravity or closed strings is not assumed microscopically: they are emergent as an effective  description.

In this way, the open/closed string correspondence supports the interpretation of gravity as an entropic force. 
Yet, many still see the closed string  side of these dualities is a well defined fundamental theory. But in our view gravity and closed strings are emergent and only present as macroscopic concept.  It just happened that we already knew about gravity before we understood it could be obtained from a microscopic theory without it. We can't resist making the analogy with a situation in which we would have developed a theory for elasticity using stress tensors in a continuous medium half a century before knowing about atoms. We probably would have been equally resistant in accepting the obvious.  Gravity and closed strings are not much different, but we just have to get used to the idea.

\subsection{Black hole horizons revisited}

 \noindent
We saved perhaps the clearest argument for the fact that gravity is an entropic force to the very last.  The first cracks in the fundamental nature of gravity appeared when Bekenstein, Hawking and others discovered the laws of black hole thermodynamics.  In fact, the thought experiment mentioned in section 3 that led Bekenstein to his entropy law is surprisingly similar to the polymer problem.
The black hole serves as the heat bath, while the particle can be thought of as the end point of the polymer that is gradually allowed to go back to its equilibrium situation.

Of course, there is no polymer in the gravity system, and there appears to be no direct contact between the particle and the black hole. But here we are ignoring the fact that one of the dimensions is emergent. In the holographic description of this same process, the particle can be thought of as being immersed in the heat bath representing the black hole. This fact is particularly obvious in the context of AdS/CFT, in which a black hole is dual to a thermal state on the boundary, while the particle is represented as a delocalized operator that is gradually being thermalized.  By the time that the particle reaches the horizon it has become part of the thermal state, just like the polymer.   This phenomenon is clearly entropic in nature, and is the consequence of a statistical process that drives the system to its state of maximal entropy. 

Upon closer inspection Bekenstein's reasoning can be used to show that gravity becomes an entropic force near the horizon, and that the equations presented in section 3 are exactly valid.
He argued that one has to choose a location slightly away from the black hole horizon at a distance of about the order of the Compton wave length, where we declare that the particle and the black hole have become one system. Let us say this location corresponds to choosing a holographic screen. The precise location of this screen can not be important, however, since there is not a natural preferred distance that one can choose. The equations should therefore not depend on small variations of this distance. 

By pulling out the particle a bit further, one changes its energy by a small amount  equal to the work done by the gravitational force. 
 If one then drops the particle in to the black hole,  the mass $M$ increases by this same additional amount. Consistency of the laws of black hole thermodynamics implies that the additional change in the Bekenstein Hawking entropy, when multiplied with the Hawking temperature $T_H$,  must be precisely equal to the work done by gravity. Hence,
 \be
F_{gravity}= T_H{\partial S_{BH}\over \partial x}.
\ee
The derivative of the entropy is defined as the response of $S_{BH}$ due to a change in the distance $x$ of the particle to the horizon.
This fact is surely known, and probably just regarded as a consistency check. But let us take it one or two steps further. 

Suppose we take the screen even further away from the horizon. The same argument applies, but we can also choose to ignore the fact that the screen has moved, and lower the particle to the location of the previous screen, the one closer to the horizon. This process would happen in the part of space behind the new screen location, and hence it should have a holographic representation on the screen.  In this system there is force in the perpendicular direction has no microscopic meaning, nor acceleration.  The coordinate $x$ perpendicular to the screen is just some scale variable associated with the holographic image of the particle. Its interpretation as an coordinate is a derived concept: this is what it means have an emergent space.

The mass is defined in terms the energy associated with the particle's holographic image, which presumably is a near thermal state. It is not exactly thermal, however, because it is still slightly away from the black hole horizon. We have pulled it out of equilibrium, just like the polymer.  One may then ask: what is the cause of the change in energy that is holographically dual to the work done when in the emergent space we gradually lower the particle towards the location of the old screen behind the new one . Of course, this can be nothing else then an entropic effect, the force is simply due to the thermalization process. We must conclude that the only microscopic explanation is that there is an emergent entropic force acting. In fact, the correspondence rules between the scale variable and energy on the one side, and the emergent coordinate $x$ the mass $m$ on the other, must be such that $F=T\nabla S$ translates in to the gravitational force.  It is straightforward to see that this indeed works and that the equations for the temperature and the entropy change are exactly as given in section 3.

The horizon is only a special location for observers that stay outside the black hole. The black hole can be arbitrarily large and the gravitational force at its horizon arbitrarily weak. Therefore, this thought experiment is  not just teaching us about black holes. It teaches us about the nature of space time, and the origin of gravity. Or more precisely, it tells us about the cause of inertia. We can do the same thought experiment for a Rindler horizon, and reach exactly the same conclusion. In this case the correspondence rules must be such that $F=T\nabla S$ translates in to the inertial force $F=ma$. Again the formulas work out as in section 3. 

\subsection{Final comments}

This brings us to a somewhat subtle and not yet fully understood aspect. Namely, the role of $\hbar$. The previous arguments make clear that near the horizon the equations are valid with $\hbar$ identified with the actual Planck constant. However, we have no direct confirmation or proof that the same is true when we go away from the horizon, or eespecially when there is no horizon present at all. 
 In fact, there are reasons to believe that the equations work slightly different there. The first is that one is not exactly at thermal equilibrium. Horizons have well defined temperatures, and clearly are in thermal equilibrium. If one assumes that the screen at an equipotential surface with $\Phi\!=\!\Phi_0$ is in equilibrium,  the entropy needed to get the Unruh temperature(\ref{unruh}) is given by the
Bekenstein-Hawking formula, including the factor $1/4$,
\be
S={c^3 \over 4G\hbar}\int_{\cal S}\!\! dA.
\ee
This value for this entropy appears to be very high, and violates the Bekenstein bound \cite{BekensteinBound} that states that a system contained in region with radius $R$ and total energy $E$ can not have an entropy larger than $ER$. The reason for this discrepancy may be that Bekenstein's argument does not hold for the holographic degrees of freedom on the screen, or because of the fact that we are far from equilibrium. 

But there may also be other ways to reconcile these statements, for example by making use of the freedom to rescale the value of $\hbar$.  This would not effect the final outcome for the force, nor the fact that it is entropic. In fact, one can even multiply $\hbar$ by a function $f(\Phi_0)$ of the Newton potential on the screen. This rescaling would affect the values of the entropy and the temperature in opposite directions: $T$ gets multiplied by a factor, while $S$ will be divided by the same factor. Since a priori we can not exclude this possibility, there is something to be understood. In fact, there are even other modifications possible, like a description that uses weighted average over many screens with different temperatures. Even then the essence of our conclusions would not change, which is the fact that gravity and inertia are entropic forces. 

Does this view of gravity lead to predictions? The statistical average should give the usual laws, hence one has to study the fluctuations in the gravitational force. Their size depends on the effective temperature, which may not be universal and depends on the effective value of $\hbar$. An interesting thought is that fluctuations may turn out to be more pronounced for weak gravitational fields between small bodies of matter. But clearly, we need a better understanding of the theory to turn this in to a prediction.  

\bigskip

It is well known that Newton was criticized by his contemporaries, especially by Hooke,  that his law of gravity acts at a distance and has no direct mechanical cause like the elastic force. Ironically, this is precisely the reason why Hooke's elastic force is nowadays not seen as fundamental, while Newton's gravitational force has maintained that status for more than three centuries. What Newton did not know, and certainly Hooke didn't,  is that the universe is holographic.  Holography is also an hypothesis, of course, and may appear just as absurd as an action at a distance. 

One of the main points of this paper is that the holographic hypothesis provides a natural mechanism for gravity to emerge. It allows direct "contact" interactions between degrees of freedom associated with one material body and another, since all bodies inside a volume can be mapped on the same holographic screen. Once this is done, the  mechanisms for Newton's gravity and Hooke's elasticity are surprisingly similar. We suspect that neither of these rivals would have been happy with this conclusion.

\newpage

\begin{flushleft}
{\bf Acknowledgements}
\end{flushleft}

This work is partly supported by Stichting FOM.  I like to thank J. de Boer, B. Chowdhuri, R. Dijkgraaf, P. McFadden, G. 't Hooft, B. Nienhuis,  J.-P. van der Schaar, and especially M. Shigemori, K. Papadodimas, and H. Verlinde for discussions and comments.


\begin{thebibliography}{100}


\bibitem{Bekenstein}
  J.~D.~Bekenstein,
  ``Black holes and entropy,''
  Phys.\ Rev.\  D {\bf 7}, 2333 (1973).


\bibitem{Bardeen}
  J.~M.~Bardeen, B.~Carter and S.~W.~Hawking,
 ``The Four laws of black hole mechanics,''
  Commun.\ Math.\ Phys.\  {\bf 31}, 161 (1973).

\bibitem{Hawking}
S .~W.~Hawking , ``Particle Creation By Black Holes,''  Commun Math. Phys. 43, 199-220, (1975)

\bibitem{Davies} 
P.~C.~W.~Davies, "Scalar particle production in Schwarzschild and Rindler metrics," J. Phys. A 8, 609 (1975)


\bibitem{Unruh}
  W.~G.~Unruh,
 ``Notes on black hole evaporation,''
  Phys.\ Rev.\  D {\bf 14}, 870 (1976).

\bibitem{Damour}
 T.~Damour, 
"Surface effects in black hole physics, in Proceedings of the Second Marcel
Grossmann Meeting on General Relativity", Ed.
R. Ruffini, North Holland, p. 587, 1982



\bibitem{Jacobson}
  T.~Jacobson,
  ``Thermodynamics of space-time: The Einstein equation of state,''
  Phys.\ Rev.\ Lett.\  {\bf 75}, 1260 (1995)


\bibitem{thooft}
  G.~'t Hooft,
  ``Dimensional reduction in quantum gravity,''
  arXiv:gr-qc/9310026.


\bibitem{susskind}
  L.~Susskind,
  ``The World As A Hologram,''
  J.\ Math.\ Phys.\  {\bf 36}, 6377 (1995)
  [arXiv:hep-th/9409089].


\bibitem{maldacena}
  J.~M.~Maldacena,
  ``The large N limit of superconformal field theories and supergravity,''
  Adv.\ Theor.\ Math.\ Phys.\  {\bf 2}, 231 (1998)
  [Int.\ J.\ Theor.\ Phys.\  {\bf 38}, 1113 (1999)]


\bibitem{Waldbook}
R.~M.~Wald,
"General Relativity,"
The University of Chicago Press, 1984


\bibitem{WaldNoether}
  R.~M.~Wald,
  ``Black hole entropy is the Noether charge,''
  Phys.\ Rev.\  D {\bf 48}, 3427 (1993)
  [arXiv:gr-qc/9307038].

\bibitem{Landscape}
  L.~Susskind,
 ``The anthropic landscape of string theory,''
  arXiv:hep-th/0302219.



\bibitem{Padma}
  T.~Padmanabhan,
 ``Thermodynamical Aspects of Gravity: New insights,'' arXiv:0911.5004 [gr-qc], and references therein.
 
 \bibitem{BekensteinBound}
  J.~D.~Bekenstein,
  ``A Universal Upper Bound On The Entropy To Energy Ratio For Bounded
  Systems,''
  Phys.\ Rev.\  D {\bf 23} (1981) 287.


\end{thebibliography}
\end{document}